# Giant Quasiparticle Bandgap Modulation in Graphene Nanoribbons Supported on Weakly Interacting Surfaces

Xueping Jiang[1,*], Neerav Kharche[1,2,*†], Paul Kohl[3], Timothy B. Boykin[4], Gerhard Klimeck[5], Mathieu Luisier[6], Pulickel M. Ajayan[7], and Saroj K.Nayak[1,8,‡]

[1] *Department of Physics, Applied Physics and Astronomy, Rensselaer Polytechnic Institute, Troy, NY 12180, USA*

[2] *Computational Center for Nanotechnology Innovations, Troy, NY, 12180, USA*

[3] *School of Chemical and Biomolecular Engineering, Georgia Institute of Technology, GA, 30332, USA*

[4] *Department of Electrical and Computer Engineering, University of Alabama in Huntsville, Huntsville, Alabama, 35899, USA*

[5] *Network for Computational Nanotechnology, Purdue University, West Lafayette, IN, 47907, USA*

[6] *Integrated Systems Laboratory, Gloriastrasse 35, ETH Zurich, Zurich, 8092, Switzerland*

[7] *Department of Mechanical Engineering & Materials Science, Rice University, 6100 Main Street, Houston, TX, 7700, USA*

[8] *School of Basic Sciences, Indian Institute of Technology Bhubaneswar, Bhubaneswar 751013, India*

**Abstract.** In general, there are two major factors affecting bandgaps in nanostructures: (i) the enhanced electron-electron interactions due to confinement and (ii) the modified self-energy of electrons due to the dielectric screening. While recent theoretical studies on graphene nanoribbons (GNRs) report on the first effect, the effect of dielectric screening from the surrounding materials such as substrates has not been thoroughly investigated. Using large-scale electronic structure calculations based on the *GW* approach, we show that when GNRs are deposited on substrates, bandgaps get strongly suppressed (by as much as 1 eV) even though the GNR-substrate interaction is weak.

[*] These authors contributed equally to this work.
[†] Present address: *Chemistry Department, Brookhaven National Laboratory, Upton, NY 11973, USA*
[‡] Electronic mail: nayaks@rpi.edu

Graphene is a two-dimensional material that exhibits extraordinary electronic, optical, thermal, and mechanical properties with potential for several technological applications.[1] Graphene confined in one dimension is called a graphene nanoribbon (GNR), which exhibits a finite bandgap.[2] Due to strong electron-electron interactions the bandgap in GNRs is known to be underestimated in an effective single-particle approach such as density functional theory (DFT).[3,4] The correct bandgap can be obtained using the many-body perturbation theory based on the Green's function ($G$) and the screened Coulomb potential ($W$) approach, i.e., the $GW$ approximation. The $GW$ approach has been employed in the past to compute bandgaps of free-standing GNRs.[4] These earlier studies report a large increase in the single-particle bandgaps due to enhanced electron-electron interactions arising from the strong one-dimensional quantum confinement in GNRs. Most of the bandgap measurements, however, are done on the substrate-supported GNRs.[2] It is well known that substrates, even the ones that weakly interact, could play an important role in modifying quasiparticle bandgap of molecules[5-7] and quantum dots.[8] The effect of substrate on the quasiparticle bandgaps of GNRs has not been thoroughly investigated so far leaving a gap in the available theoretical predictions and the experimental measurements.

In this paper, we investigate the effect of substrates on the quasiparticle bandgap of GNRs using the many-body perturbation theory in the $GW$ approximation, a state-of-the-art electronic structure method. We find that the bandgaps of GNRs drastically reduce when they are deposited on a substrate. For example, a free-standing armchair GNR (AGNR) of width 0.68 nm has a bandgap of 1.72 eV, but when deposited on a weakly interacting hexagonal boron nitride (hBN) substrate the bandgap is reduced to 0.93 eV. The large bandgap modulation in GNRs due to the underlying substrate is related to the modification of the self-energy of electrons by the non-local dielectric screening of electron-electron interactions in GNRs due to the polarization of the substrate surface. The bandgap reduction is strongly dependent on the edge geometry due to the peculiar dependence of the electronic confinement on the edge geometry of GNRs. Furthermore, as expected from this analysis the same free standing GNRs, when deposited on other weakly interacting substrates with larger dielectric constants such as graphite, have smaller bandgaps compared to that on hBN substrate.

Graphite and hBN substrates are chosen for the following reasons: minimal lattice mismatch between atomic structures of the GNR and the substrate, van der Waals interaction between them, and absence of any dangling bonds on the substrate surface[9] so that we can unequivocally relate electronic structure modulation to the dielectric screening. The findings presented here are also applicable to GNRs deposited on the other weakly interacting substrates such as $SiO_2$ and SiC.[10]

The schematics of the simulated GNRs deposited on the hBN substrate are shown in Fig. 1. The edges of GNRs are passivated with hydrogen. The GNRs deposited on hBN are modeled using the repeated slab approach, where the slabs periodic in the *xy* plane are separated by a large enough vacuum region ($\approx 12$ Å) so that their interaction with the periodic images along the *z* direction is negligible. The lateral distance between edges of GNR and edges of its periodic images is $\approx 8$ Å, which insures negligible interaction between the periodic images of GNR along the *x* direction. The simulated geometries of the free-standing GNRs are identical and the same vacuum regions along *x* and *z* directions are used. The hBN-supported GNRs are oriented according to the AB Bernal stacking with respect to the underlying hBN substrate as shown in Figs. 1(b) and (c). In the following discussion, the GNRs are labeled by the number of carbon atoms along the width direction.

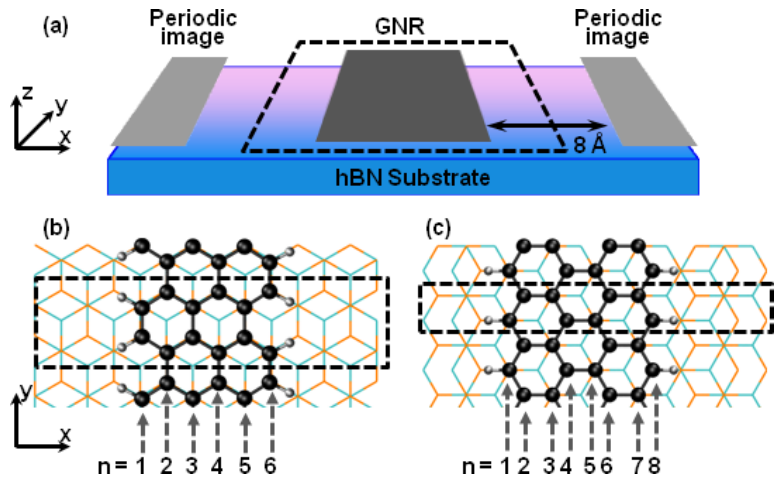

**Fig. 1. Simulated structures:** (a) Schematic of a GNR deposited on hBN substrate and its periodic images along lateral (x) directions. The simulation domain projected onto the xy plane is depicted by dashed lines and is periodic in the x and y directions. GNR is effectively periodic along the y direction due to the 8 Å separation from its periodic images. Atomistic schematics of hBN-supported (b) AGNR-6 and (c) ZGNR-8. GNRs are shown in the ball-and-stick representation and two monolayers of hBN substrate are shown in the wireframe representation. The unit cells are depicted by the dashed rectangles.

The electronic structure calculations are performed using the local density approximation (LDA) as implemented in the ABINIT code.[11] The Trouiller-Martins norm-conserving pseudopotentials[12] and the Teter-Pade parameterization for the exchange-correlation functional[13] are used. The quasiparticle bandgaps are calculated using the many-body perturbation theory within the $G_0W_0$ approximation and the screening is calculated using the plasmon-pole model[14]. Two monolayers of the semi-infinite hBN/graphite substrate are included in the simulation domain to ensure that the *GW* bandgap is converged. The 2D Brillouin zone of the substrate-supported GNRs is sampled using Monkhorst-Pack *k*-point grids of 10×4×1 for AGNRs and 20×4×1 for ZGNRs. The 1D Brillouin zone of the free-standing GNRs is sampled using Monkhorst-Pack grids of 10×1×1 for AGNRs and 20×1×1 for ZGNRs. Wave functions are expanded in plane waves with an energy cutoff of 18 Ha. The results converge very well based on the above data. The coulomb cutoff[15]

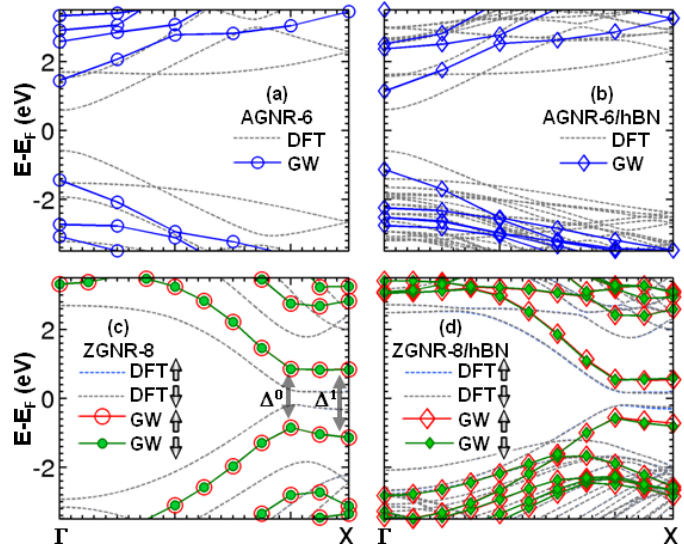

**Fig. 2. DFT and GW band structures of free-standing and hBN-supported GNRs:** (a) Free-standing and (b) hBN-supported AGNR-6. (c) Free-standing and (d) hBN-supported ZGNR-8. The additional bands in (b) and (d) are contributed by the hBN substrate. In (c), the direct bandgap and the energy gap at the zone-boundary are denoted by $\Delta^0$ and $\Delta^1$ respectively.

technique is used to minimize the spurious interactions with periodic replicas of the system.

The geometry optimizations are carried out using the Vienna Ab Initio Simulation Package (VASP)[16], which provides a well-tested implementation of van der Waals (vdW) interactions[17]. The PAW[18] pseudopotentials, the Perdew-Burke-Ernzerhof (PBE) exchange-correlation functional in the generalized gradient approximation (GGA)[19], and the DFT-D2 method of Grimme[20] are used. The same values of energy cutoff, vacuum region and $k$-point grid as in the ABINIT calculations are used in VASP calculations. All of the structures were relaxed using a conjugate gradient algorithm with the atomic force tolerance of 0.05 eV/Å and the total energy tolerance of $10^{-4}$ eV. The optimized geometries calculated using VASP and ABINIT without including vdW interactions are found to agree well with each other.

The DFT and *GW* bandstructures of AGNR-6 (width $w$=0.8 nm) in the free-standing configuration are shown in Fig. 2(a). Similar to earlier studies, the quasiparticle *GW* bandgap of the free-standing AGNR-6 is found to be significantly higher compared to the DFT bandgap. The *GW* correction to the DFT bandgap of the free standing AGNR-6 is 1.72 eV. Such a large *GW* correction is attributed to (i) the enhanced electron-electron interaction due to the strong one-dimensional quantum confinement in GNRs and (ii) a weak screening of Coulomb interaction in GNRs by the surrounding vacuum.[4]

Fig. 2(b) shows the DFT and *GW* band structures of AGNR-6 when deposited on the hBN substrate. AGNR binds weakly with the hBN substrate by van der Waals forces such that the distance between them is ≈ 3.2 Å. Near the Fermi energy, the DFT band structures of AGNR-6 in both the free-standing and substrate-supported configurations are identical. This is because of the fact that there is no hybridization between the GNR and hBN wave functions in this energy range. As a result, in the DFT formalism, which does not include the long-range correlation effects, AGNR-6 and hBN substrate are essentially electronically isolated systems. In contrast, in the *GW* calculation, which includes the long-range correlation effects, the bandgap is significantly reduced when AGNR-6 is deposited on the hBN substrate. The *GW* bandgaps of free-standing and substrate-supported AGNR-6 are 2.89 eV and 2.27 eV respectively, which corresponds to 0.62 eV bandgap reduction. AGNRs belonging to all three classes (i.e. $N$=3$p$, 3$p$+1, and 3$p$+2, where $N$ denotes the number of carbon chains along the width and $p$ is an integer) exhibit similar bandgap reduction when they are supported on the hBN substrate.

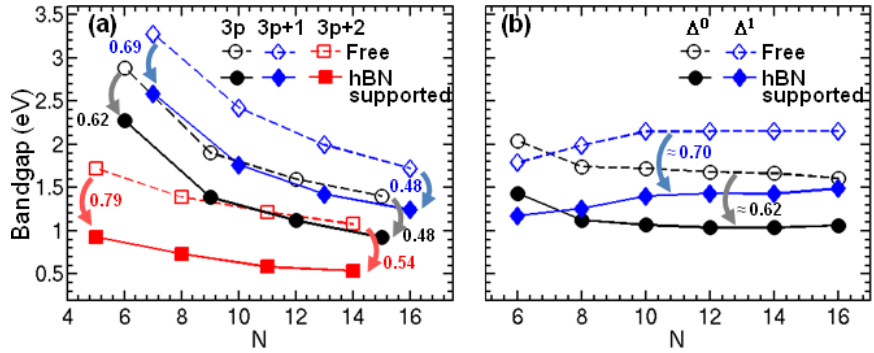

**Fig. 3. Bandgap variation with width:** (a) Quasiparticle bandgaps of the three families of AGNRs. (b) Quasiparticle gaps $\Delta^0$ and $\Delta^1$ of the free-standing and hBN-supported. The open symbols correspond to the AGNRs in the free-standing configuration while the solid symbols correspond to the hBN-supported AGNRs.

Figs. 2(c) and (d) show the DFT and *GW* band structures of ZGNR-8 in the free-standing and hBN-supported configurations respectively. Both the free-standing and hBN supported ZGNR-8 have antiferromagnetic ground state[3,4,21] with almost identical DFT band structures while the *GW* band structures show a substrate polarization induced bandgap reduction similar to AGNRs.

The *GW* bandgap reduction in the hBN-supported GNRs is a result of more effective screening of the Coulomb interaction in GNRs by the hBN substrate compared to the screening by the vacuum in free-standing GNRs. The polarization of the hBN surface reduces the screened Coulomb potential ($W$) in GNRs, which in turn reduces the *GW* bandgap. Similar bandgap reductions have been reported in experiments and *GW* calculations on zero-dimensional systems such as molecules adsorbed on dielectric surfaces,[5,7,22] quantum dots[8] embedded in dielectric matrices, and two-dimensional systems such as heterostructures of hydrogenated graphene and hBN[23].

The substrate surface polarization is dependent on the electronic confinement in GNRs, which gives rise to a peculiar dependence of the bandgap reduction on the width and edge geometry of GNRs. The quasiparticle bandgaps of the free-standing and hBN-supported AGNRs with the widths ranging from 0.68 nm to 1.8 nm are shown in Fig. 3(a). The quasiparticle bandgaps of all three families of AGNRs are suppressed when they are deposited on the substrate. The substrate induced bandgap reductions in all three families of AGNRs show a clear width dependence such that the wider AGNRs exhibit smaller bandgap reduction compared to the narrower AGNRs. The width dependence arises from the fact that the charge distributions of the states at the valence band maximum and the conduction band minimum are more delocalized in the wider AGNRs compared to the narrower AGNRs. The substrate polarization induced screening is less effective for the delocalized charge distribution compared to the localized charge distribution[5].

Fig. 3(b) shows the quasiparticle bandgaps $\Delta^0$ (depicted in Fig. 2(c)) of the free-standing and hBN-supported ZGNRs of the widths ranging from 0.81 nm to 1.87 nm. In contrast to the AGNRs, the substrate induced bandgap reduction in ZGNRs does not show any width dependence but fluctuates around 0.62 eV. Unlike AGNRs where the wavefunctions of the conduction band minimum and valence band maximum states are spread across the width, in ZGNRs the wavefunctions of these $\Delta^0$ states are localized at the edges[4]. Due to the so-called edge-state nature, the charge distributions of the valence band and the conduction band $\Delta^0$ states are almost independent of the width of ZGNRs. Consequently, the substrate polarization induced screening and the resulting bandgap reduction is virtually independent of the width.

The energy gap at the Brillouin zone boundary ($\Delta^1$) of the free-standing and hBN-supported ZGNRs is also shown in Fig. 3(a). The wavefunctions of the $\Delta^1$ states are more localized at the edges compared to the $\Delta^0$ states. Similar to the earlier study, the more localized $\Delta^1$ states are found to have larger $GW$ self-energy correction compared to the less localized $\Delta^0$ states.[4] Compared to $\Delta^0$ states, the quasiparticle energies of the more localized $\Delta^1$ states exhibit stronger renormalization due to the substrate polarization. As a result, the $\Delta^1$ energy gap exhibits larger substrate induced bandgap reduction compared to the $\Delta^0$ energy gap. Thus the substrate polarization induced bandgap reduction strongly depends on the edge geometry of GNRs due to the peculiar electronic confinement in these nanostructures.

We also performed similar electronic structure calculations on GNRs deposited on the graphite substrate. Graphite, due to its semimetallic nature screens the Coulomb interaction in GNRs more effectively compared to hBN resulting in stronger bandgap suppression. For example, the $GW$ bandgap of AGNR-6 is reduced by 1.2 eV when it is deposited on graphite substrate compared to 0.62 eV bandgap reduction when deposited on hBN substrate. Thus the bandgap reduction is strongly dependent on the dielectric properties of the substrate even in the weakly interacting regime. Compared to GNRs, small organic molecules such as benzene and pentacene exhibit much larger bandgap reduction (benzene: 3.20 eV and pentacene: 2.36 eV) when deposited on graphite[5]. This is because of the fact that GNR screens electron-electron interactions more effectively compared to small molecules. As a result the additional screening from substrate has less effect on bandgaps in GNRs compared to that in small molecules.

The results presented here have important implications for the design of carbon-based electronics. For example, the bandgap adversely affects the performance of GNRs when used as interconnects[24]. The study suggests that the performance of GNR interconnects can be improved by surrounding GNRs with high-κ dielectrics, which screen electron-electron interactions in GNRs thereby reducing their bandgaps. On the contrary, when GNRs are used as a channel material in field-effect transistors (FETs), large bandgaps are more desirable.[24] In this case low-κ dielectrics are more suitable to minimize the bandgap reduction due to the dielectric screening. Interestingly, this prescription is opposite to the current trend in silicon-based nanoelectronics where low-κ dielectrics are preferred for interconnects while high-κ dielectrics are preferred for FETs.[24]

Recent experimental studies observed bandgap reduction in substrate supported AGNRs[25], which supports our theoretical predictions. Earlier theoretical work employed classical image-charge model to estimate the bandgap reduction[25,26]. The present work reports results from first principle simulations without any fitting parameter.

To summarize, the many-body perturbation theory based on the *GW* approach shows that substrates play an important role in controlling bandgaps in GNRs. The bandgaps of free standing GNRs are reduced drastically, sometimes by as much as 1 eV or more in spite of weak van der Waals interactions between GNR and the underlying substrate. While strong electron-electron interactions due to quantum confinement open up large bandgaps in GNRs, the underlying substrates can suppress electron-electron interactions thereby reducing GNR bandgaps. The results further suggest that contrary to usual prescription in silicon-based electronics, high-κ materials could be better for GNR-based interconnects while low-κ materials could more suitable for FETs. The results presented here could be exploited in making new electronic materials beyond carbon nanostructures where long-range polarization effect could be used to tune electronic and magnetic properties of low dimensional structures.

**Acknowledgements:** This work is supported partly by the Interconnect Focus Center (IFC) and the Materials Structures and Devices (MSD) focus center funded by the MARCO program of SRC and State of New York, NSF PetaApps grant number OCI-0749140, NSF Columbia, and an anonymous gift from Rensselaer. Computing resources of the Computational Center for Nanotechnology Innovations at Rensselaer partly funded by State of New York and of nanoHUB.org funded by the National Science Foundation have been used for this work. SKN thanks Professor Mike Payne and Cavendish laboratory for their hospitality.

**References**

[1] F. Schedin, A. K. Geim, S. V. Morozov, E. W. Hill, P. Blake, M. I. Katsnelson, and K. S. Novoselov, Nat. Mater. **6** (9), 652 (2007); A. A. Balandin, S. Ghosh, W. Z. Bao, I. Calizo, D. Teweldebrhan, F. Miao, and C. N. Lau, Nano Lett. **8** (3), 902 (2008); F. Schwierz, Nat. Nanotechnol. **5** (7), 487 (2010); A. K. Geim and K. S. Novoselov, Nat. Mater. **6** (3), 183 (2007); K. K. Saha, M. Drndic, and B. K. Nikolic, Nano Lett. **12** (1), 50 (2012); H. W. C. Postma, Nano Lett. **10** (2), 420 (2010).

[2] Melinda Y. Han, Barbaros Özyilmaz, Yuanbo Zhang, and Philip Kim, Physical Review Letters **98** (20), 206805 (2007); X. R. Wang, Y. J. Ouyang, L. Y. Jiao, H. L. Wang, L. M. Xie, J. Wu, J. Guo, and H. J. Dai, Nat. Nanotechnol. **6** (9), 563 (2011); J. M. Cai, P. Ruffieux, R. Jaafar, M. Bieri, T. Braun, S. Blankenburg, M. Muoth, A. P. Seitsonen, M. Saleh, X. L. Feng, K. Mullen, and R. Fasel, Nature **466** (7305), 470 (2010); T. Shimizu, J. Haruyama, D. C. Marcano, D. V. Kosinkin, J. M. Tour, K. Hirose, and K. Suenaga, Nat. Nanotechnol. **6** (1), 45 (2011); S. Linden, D. Zhong, A. Timmer, N. Aghdassi, J. H. Franke, H. Zhang, X. Feng, K. Mullen, H. Fuchs, L. Chi, and H. Zacharias, Physical Review Letters **108** (21), 216801 (2012).

[3] Young-Woo Son, Marvin L. Cohen, and Steven G. Louie, Physical Review Letters **97** (21), 216803 (2006).

[4] Li Yang, Cheol-Hwan Park, Young-Woo Son, Marvin L. Cohen, and Steven G. Louie, Physical Review Letters **99** (18), 186801 (2007).

[5] J. B. Neaton, M. S. Hybertsen, and S. G. Louie, Physical Review Letters **97** (21), 216405 (2006).

[6] J. M. Garcia-Lastra, C. Rostgaard, A. Rubio, and K. S. Thygesen, Physical Review B **80** (24), 245427 (2009).

[7] Y. Li, D. Y. Lu, and G. Galli, Journal of Chemical Theory and Computation **5** (4), 881 (2009).

[8] A. Franceschetti and A. Zunger, Appl. Phys. Lett. **76** (13), 1731 (2000).

[9] C. R. Dean, A. F. Young, I. Meric, C. Lee, L. Wang, S. Sorgenfrei, K. Watanabe, T. Taniguchi, P. Kim, K. L. Shepard, and J. Hone, Nat. Nanotechnol. **5** (10), 722 (2010); J. M. Xue, J. Sanchez-Yamagishi, D. Bulmash, P. Jacquod, A. Deshpande, K. Watanabe, T. Taniguchi, P. Jarillo-Herrero, and B. J. Leroy, Nat. Mater. **10** (4), 282 (2011).

[10] M. Z. Hossain, Appl. Phys. Lett. **99** (18), 183103 (2011); D. M. Zhang, Z. Li, J. F. Zhong, L. Miao, and J. J. Jiang, Nanotechnology **22** (26), 265702 (2011).

[11] X. Gonze, B. Amadon, P. M. Anglade, J. M. Beuken, F. Bottin, P. Boulanger, F. Bruneval, D. Caliste, R. Caracas, M. Cote, T. Deutsch, L. Genovese, P. Ghosez, M. Giantomassi, S. Goedecker, D. R. Hamann, P. Hermet, F. Jollet, G. Jomard, S. Leroux, M. Mancini, S. Mazevet, M. J. T. Oliveira, G. Onida, Y. Pouillon, T. Rangel, G. M. Rignanese, D. Sangalli, R. Shaltaf, M. Torrent, M. J. Verstraete, G. Zerah, and J. W. Zwanziger, Comput. Phys. Commun. **180** (12), 2582 (2009).

[12] N. Troullier and J. L. Martins, Physical Review B **43** (3), 1993 (1991).

[13] S. Goedecker, M. Teter, and J. Hutter, Physical Review B **54** (3), 1703 (1996).

[14] M. S. Hybertsen and S. G. Louie, Physical Review B **34** (8), 5390 (1986).

[15] S. Ismail-Beigi, Physical Review B **73** (23) (2006).

[16] G. Kresse and J. Furthmuller, Comput. Mater. Sci. **6** (1), 15 (1996).

[17] T. Bucko, J. Hafner, S. Lebegue, and J. G. Angyan, J. Phys. Chem. A **114** (43), 11814 (2010).

[18] G. Kresse and D. Joubert, Phys. Rev. B **59** (3), 1758 (1999).

[19] J. P. Perdew, K. Burke, and M. Ernzerhof, Phys. Rev. Lett. **78** (7), 1396 (1997).

[20] S. Grimme, J. Comput. Chem. **27** (15), 1787 (2006).

[21] M. Fujita, K. Wakabayashi, K. Nakada, and K. Kusakabe, J. Phys. Soc. Jpn. **65** (7), 1920 (1996); N. Kharche, Y. Zhou, K. P. O'Brien, S. Kar, and S. K. Nayak, ACS Nano **5** (8), 6096 (2011).

[22] K. S. Thygesen and A. Rubio, Physical Review Letters **102** (4), 046802 (2009).

[23] N. Kharche and S. K. Nayak, Nano Lett. **11** (12), 5274 (2011).

[24] International Technology Roadmap for Semiconductors (ITRS), (2011).

[25] P. Ruffieux, J. M. Cai, N. C. Plumb, L. Patthey, D. Prezzi, A. Ferretti, E. Molinari, X. L. Feng, K. Mullen, C. A. Pignedoli, and R. Fasel, Acs Nano **6** (8), 6930 (2012).

[26] Liangbo Liang and Vincent Meunier, Physical Review B **86** (19), 195404 (2012).